\documentclass[12pt]{article}

\usepackage{graphics,amssymb,epsfig,float}
\usepackage[usenames,dvips]{color}
\usepackage{graphicx}% Include figure files
\usepackage{epsfig}% Include figure files
\usepackage{rotating}% Include figure files
\usepackage{dcolumn}% Align table columns on decimal point
\usepackage{bm}% bold math
\usepackage{cite}
\usepackage{amsmath}
\usepackage{array}

\makeatletter

\@addtoreset{equation}{section}
\makeatother

\def\lsim{\;\raise0.3ex\hbox{$<$\kern-0.75em\raise-1.1ex\hbox{$\sim$}}\;}
\def\gsim{\;\raise0.3ex\hbox{$>$\kern-0.75em\raise-1.1ex\hbox{$\sim$}}\;}
\def\beq{\begin{equation}}   \def\eeq{\end{equation}}
\def\ba{\begin{array}}       \def\ea{\end{array}}
\def\bea{\begin{eqnarray}}   \def\eea{\end{eqnarray}}
\def\nn{\nonumber}
\def\nl{\newline}

\begin{document}

\begin{titlepage}
\begin{flushright}
LPT Orsay 14-38 \\
PCCF RI 14-06\\
\end{flushright}

%\centerline{\bf \today}

\begin{center}
\vspace{1cm}
{\Large\bf NMSSM with a singlino LSP: possible challenges for searches for supersymmetry at the LHC} \\
\vspace{2cm}

{\bf{Ulrich Ellwanger$^a$ and Ana M. Teixeira$^b$}}\\
\vspace{1cm}
\it  $^a$ LPT, UMR 8627, CNRS, Universit\'e de Paris--Sud, 91405 Orsay,
France, and \\
\it School of Physics and Astronomy, University of Southampton,\\
\it Highfield, Southampton SO17 1BJ, UK\\
\it $^b$ Laboratoire de Physique Corpusculaire, CNRS/IN2P3 - UMR 6533\\
Campus des C\'ezeaux, 24 Av. des Landais, F-63171 Aubi\`ere, France 

\end{center}
\vspace{2cm}

\begin{abstract}
A light singlino in the NMSSM can reduce considerably the missing transverse
energy at the end of sparticle decay cascades; instead, light NMSSM-specific
Higgs bosons can be produced. Such scenarios can be consistent with present
constraints from the LHC with all sparticle masses below $\sim 1$~TeV. We discuss
search strategies, which do not rely on missing transverse
energy, for such scenarios at the next run of the LHC near 14~TeV.

\end{abstract}

\end{titlepage}

\section{Introduction}

One of the main tasks of the LHC was and will be the search for
supersymmetric (SUSY) particles. The largest production cross sections
are expected for gluinos ($\tilde{g}$) and squarks ($\tilde{q}$) of the
first generation. After the first run of the LHC at a center of mass (c.m.)
energy of mostly 8~TeV, no significant excesses have been observed in corresponding
search channels \cite{TheATLAScollaboration:2013fha, Chatrchyan:2014lfa}
(see \cite{Melzer-Pellmann:2014eta} for a recent summary).

The absence of excess events can be interpreted in terms of lower bounds on gluino
and squark masses, once assumptions on their decay cascades are made.
These depend on the masses and couplings of many other SUSY particles, at least
on the mass of the lightest SUSY particle (LSP). Within simplified models
(assuming simple 1-step decay cascades) or the Minimal SUSY extension of the Standard
Model (MSSM), lower bounds on gluino and squark masses are typically in
the 1.2-2~TeV range, and $\sim$ 1.7~TeV if gluino and squark masses are assumed to
be similar \cite{TheATLAScollaboration:2013fha, Chatrchyan:2014lfa}.
Although these constraints do not rule out the MSSM, they eliminate a
significant part of its ``natural'' parameter space \cite{Feng:2013pwa}.

However, the MSSM is not the only SUSY extension of the Standard Model (SM) which
alleviates the hierarchy problem, provides an acceptable dark matter candidate and
leads to Grand Unification of the running gauge couplings. In the present
paper we consider the Next-to-Minimal SUSY extension of the Standard
Model (NMSSM) \cite{Ellwanger:2009dp}, where the coupling of the two Higgs
doublets of the MSSM to an additional gauge singlet field $S$ renders more
natural a value
of $\sim 126$~GeV of the SM-like Higgs boson
\cite{Hall:2011aa,Ellwanger:2011aa,Arvanitaki:2011ck,King:2012is,
Kang:2012sy,Cao:2012fz}, while preserving the attractive features of the MSSM.
Besides the
Higgs sector, the NMSSM differs from the MSSM through the presence of an
additional neutralino (the singlino, the fermionic component of the singlet
superfield). The singlino can be the LSP, which can modify considerably the
SUSY particle decay cascades \cite{Ellwanger:1997jj,Ellwanger:1998vi, Dedes:2000jp,Choi:2004zx,Cheung:2008rh,Stal:2011cz,Cerdeno:2013qta,
Das:2012rr,Das:2013ta}.

The strongest constraints from searches for gluinos and squarks
of the first generation originate
from channels where one looks for events with jets with large transverse
momentum $p_T$ and missing transverse energy $E_T^\mathrm{miss}$
\cite{TheATLAScollaboration:2013fha, Chatrchyan:2014lfa}. 
The $E_T^\mathrm{miss}$ is due to having all SUSY decay cascades ending
in a stable LSP (under the assumption of R-parity conservation), which
escapes detection (if neutral, as required for dark matter).

In the present paper we point out that a singlino-like LSP in the NMSSM can
reduce significantly the missing transverse energy at the end of SUSY 
particle decay cascades. This is due to the kinematics of the last process in
a SUSY particle decay chain, NLSP $\to$ LSP + $X$, where NLSP denotes the
Next-to-LSP, and $X$ a particle (e.g. a Higgs boson) decaying into visible
components of the SM. For a light LSP, if the mass of
$X$ is close to the mass of the NLSP, little energy and momentum are transferred
from the NLSP to the LSP; most of the energy is transferred to $X$. Correspondingly,
the LSP in the final state leads to little $E_T^\mathrm{miss}$, whereas large
$E_T^\mathrm{miss}$ is one of
the relevant search criteria for SUSY particles in general.

%UE
The possibility to discover squarks and gluinos without relying on
$E_T^\mathrm{miss}$, but on leptons, has been studied earlier in
\cite{Baer:2008kc,Baer:2009dn,Lisanti:2011tm}. \cite{Lisanti:2011tm}
discuss decays of an NLSP into a scalar (decaying visibly into SM particles)
and the LSP, referring to the NMSSM without,
however, considering the particular kinematic configurations analysed below.

A scenario similar to the one discussed here has been named ``Stealth
Supersymmetry'' \cite{Fan:2011yu, Fan:2012jf}. There, however,
a complete ``stealth sector'' is added to the MSSM in order to obtain
the above kinematic configuration of the NLSP decay. 

An extensive survey of present constraints on gluinos from searches,
including several without relying on $E_T^\mathrm{miss}$, is given in
\cite{Evans:2013jna}. Among the scenarios analysed in
\cite{Evans:2013jna} are so-called ``minimal Hidden Valley'' models.
These are similar to the ones considered here after
replacing the extra singlet scalar and its fermionic superpartner
\cite{Evans:2013jna} by the corresponding states of the
NMSSM (and the NLSP higgsino by a bino-like NLSP). It was already found in
\cite{Evans:2013jna} that the kinematic configuration discussed above
leads to the weakest constraints.

The present scenario is opposite to the one of ``compressed supersymmetry'' 
\cite{Martin:2007gf,LeCompte:2011cn,LeCompte:2011fh,Belanger:2012mk,
Dreiner:2012gx,Bhattacherjee:2013wna} where the masses of the NLSP
and the LSP are assumed to be similar, and little energy is transferred to
$X$. Then jets (or leptons) with large $p_T$ would be rare. Moreover, unless
a hard jet is emitted from the initial state (``monojet''), the
$E_T^\mathrm{miss}$ due to two LSPs emitted from two SUSY particles
back-to-back in the transverse plane tend to cancel.

In the MSSM, the kinematic configuration considered here
cannot play a major r\^ole:
%UEend
A light LSP (with a mass of a few GeV) can only be bino-like, since winos
or higgsinos would have charged partners with similar masses, already
ruled out by LEP. All squarks -- appearing also in gluino decays
-- have hypercharge and hence couple to the bino. If the LSP is a very
light bino, squarks will in general decay directly into the bino, and hardly pass
through an NLSP (e.g. a heavier neutralino or chargino) and a state $X$
(a Higgs, $Z$ or $W$ boson). Thus only a fraction of cascade
decays leads to a reduction of $E_T^\mathrm{miss}$, so that the interpretation
of the absence of signal events in terms of lower bounds on SUSY particle
masses remains practically unchanged.

On the other hand, in the NMSSM the bino can be the NLSP, the singlino
a light LSP, and $X$ a priori a Higgs, a $Z$ or even a $W$ boson (if the NLSP is
a chargino). Then the decays of $X$ can still give rise to missing energy in
the form of neutrinos; this is the case for the decays of the $W$ and $Z$
bosons, and also for the SM-like Higgs (when it decays via $WW^*$ or $ZZ^*$).

However, in the NMSSM additional Higgs bosons exist, which can be
lighter than the $Z$ boson and are not excluded by LEP
due to small couplings to $ZZ$. A lighter CP-even Higgs boson $H_1$
with a mass below $M_Z$ would have very small decay rates into $WW^*$, but
decay dominantly into $b\bar{b}$ and, to some extent, into $\tau^+\tau^-$.
Although the latter decays can also give rise to some $E_T^\mathrm{miss}$,
the scenario NLSP $\to$ LSP + $H_1$ with $M_{H_1} \lsim M_\mathrm{NLSP} < M_Z$
would be the most difficult one with respect to signatures based 
on~$E_T^\mathrm{miss}$. (Subsequently we denote scenarios with as little
$E_T^\mathrm{miss}$ as possible as ``worst case''.)

In the case of squark/gluino pair production, some $E_T^\mathrm{miss}$ can
also originate from $W$, $Z$ and/or Higgs decays which appear during decay cascades
involving charginos and/or heavier neutralinos. Again, a ``worst case''
scenario would be one where this does not happen if, for instance, the
chargino and heavier neutralino masses are close to (or above) the squark masses.

In the present paper we concentrate on such ``worst case'' scenarios:
First, we present the properties of points in the NMSSM
which are not excluded by present SUSY searches although all sparticle
masses are below $\sim 1$~TeV. Second, we propose search strategies for
these difficult scenarios, putting forward an analysis of events
at the LHC near 14~TeV c.m. energy, based on the decay products of two
$H_1$ bosons in the $b\bar{b}\tau^+\tau^-$ + jets final state.
Our simulations indicate that, for not excessively heavy squarks and gluinos
(i.e. a not too small production cross section), a signal can be visible
above Standard Model backgrounds.

In the next section we discuss in detail scenarios within the
general NMSSM, in which $E_T^\mathrm{miss}$ is reduced for kinematic
reasons. Results of event simulations of such a benchmark point are
discussed, which explain the reduced sensitivity of present SUSY searches
to such a scenario. We also discuss simplified models with varying LSP and $H_1$
masses, and the corresponding reduction of signal events. In Section~3
we attempt to extract signals for $H_1$ pair production at the LHC with
14~TeV c.m. energy, with dedicated cuts which do not rely on $E_T^\mathrm{miss}$.
Instead,
we attempt to identify $b$-jets and $\tau$-leptons from boosted $H_1$
bosons with the help of a jet reconstruction with a small jet cone radius
$R=0.15$. Section~4 contains a summary and conclusions.

\section{``Missing'' missing energy in the NMSSM}

Given a possible last step in a SUSY particle decay chain NLSP $\to$ LSP + $X$
in the limit of a narrow phase space, $M_{\mathrm{NLSP}}-(M_{\mathrm{LSP}}+M_X)
\ll M_{\mathrm{NLSP}}$, the energy (momentum) transferred from the NLSP to the LSP in the laboratory frame
is proportional to the ratio of masses:
\beq\label{eq:2.1}
{\frac{E_{\mathrm{LSP}}}{E_{\mathrm{NLSP}}} \simeq \frac{M_{\mathrm{LSP}}}
{M_{\mathrm{NLSP}}}}\; .
\eeq
Hence, if the LSP is light, little (missing transverse) energy is transferred
to the LSP; the transverse energy is carried away by $X$. The effect is
the more important the narrower the phase space is.
As explained in the
Introduction, such a scenario is difficult to realise in the MSSM where
such a light LSP must be bino-like.

The particle content of the NMSSM differs from that of the MSSM by an additional
singlino-like neutralino $\tilde{S}$, and additional singlet-like CP-even and
CP-odd Higgs bosons \cite{Ellwanger:2009dp}. Notably the NMSSM spectrum contains
three CP-even Higgs bosons $H_i$, $i=1,2,3$ (ordered in mass).
The singlino-like neutralino
can be the LSP with a bino-like NLSP (as occurs for
the regions of parameter space considered below). Then the above
scenario of little $E_T^\mathrm{miss}$ being transferred to the LSP
can be realised with a singlet-like CP-even Higgs boson $H_1$
playing the r\^ole of $X$, whose subsequent decays give rise to little
invisible transverse energy in the form of neutrinos. Typical values for the
masses would be a few GeV for the singlino-like neutralino $\tilde{S}$, a
bino-like NLSP with a mass $M_\mathrm{bino}$ just below $M_Z$, and 
$M_{H_1}$ just below $M_\mathrm{bino} - M_{\tilde{S}}$. Note that, due
to its reduced coupling to the $Z$ boson, such a light $H_1$ can still
be compatible with constraints from Higgs searches at LEP 
\cite{Schael:2006cr}.

In the simplest $\mathbb Z_3$ invariant realisation of the NMSSM, the
diagonal elements of the mass matrices for the (pure) singlet-like states
$\tilde{S}$, $H_S$ and $A_S$ satisfy \cite{Das:2012rr}
\beq\label{eq:2.2}
M_{\tilde{S}}^2 \sim M_{H_S}^2 + \frac{1}{3}M_{A_S}^2\; ,
\eeq
which forbids $M_{H_S} \gg M_{\tilde{S}}$ and hence $M_{H_1} \gg
M_{\mathrm{LSP}}$ unless soft SUSY breaking trilinear
couplings are in the multi-TeV range, in which case there can be strong
deviations from the equality of Eq.~(\ref{eq:2.2}) for the mass eigenstates
(after diagonalization of the mass matrices).

However, $M_{H_S} \gg M_{\tilde{S}}$ is possible in the
presence of $\mathbb Z_3$ violating terms like a soft SUSY breaking
tadpole term $\xi_S S$, and/or a holomorphic soft SUSY breaking mass
term $\frac{1}{2}{m'_S}^2 S^2 + \mathrm{h.\ c.}$. Such terms are
generated automatically in gauge mediated supersymmetry breaking (GMSB), if
the singlet superfield has couplings to the messenger fields
\cite{Ellwanger:2008py}.

Hence we consider in the following a general NMSSM, still with a
$\mathbb Z_3$ invariant superpotential
\beq\label{eq:2.3}
W_\mathrm{NMSSM} = \lambda \hat S \hat H_u\cdot \hat H_d +
\frac{\kappa}{3} 
\hat S^3 + \dots\; .
\eeq
In the above hatted letters denote superfields, and the ellipses denote the
MSSM-like Yukawa couplings of $\hat H_u$ and $\hat H_d$ to the
quark and lepton superfields. We allow for the following
NMSSM specific soft SUSY breaking terms
\beq\label{eq:2.4}
-{\cal L}_\mathrm{NMSSM}^\mathrm{soft} = m_S^2 |S|^2 +\left(\lambda
A_\lambda H_u H_d S + \xi_S S +\frac{1}{2}{m'_S}^2 S^2
+  \frac{1}{3}\kappa A_\kappa S^3\right) +  \mathrm{h.\ c.}\; .
\eeq
As can be seen from Eq.~(\ref{eq:2.3}), a vacuum expectation value
$\left<S\right>$ generates an effective $\mu_\mathrm{eff}$ term
$\mu_\mathrm{eff}=\lambda \left<S\right>$, which has to be larger
than $\sim 100$~GeV for the charged higgsinos
to satisfy bounds from LEP. Given the diagonal singlino mass term 
$M_{\tilde{S}} = 2\kappa\left<S\right>$, a singlino mass of a
few GeV is obtained for $\kappa$ about two orders of magnitude smaller
than $\lambda$.

For completeness we comment on the possibilities to obtain consistent
properties of dark matter in such a scenario. Within GMSB,
a gravitino can be lighter than $\tilde{S}$ which
would thus not be the ``true'' LSP, but decay radiatively into a gravitino
and a photon
(through a small photino component from a non-vanishing mixing with the bino/wino).
However, the singlino life time would be so large that this decay would
occur outside the detectors and have no impact on our subsequent
analyses. 
(On the other hand, the singlino life time
should not exceed $\sim 100$~s in order not to spoil nucleosynthesis
unless the NLSP density is diluted through entropy production.)
Alternatively, the singlino relic density can be reduced to comply with the
observed dark matter relic density through the
exchange of a CP-odd Higgs state $A_S$ in the s-channel, provided
$M_{A_S} \sim 2 M_{\tilde{S}}$. We have checked that this is indeed
possible, and the benchmark point given below has this property.

Returning to the issue of $E_T^\mathrm{miss}$, its suppression is
maximised if no neutrinos from $Z/W$ decays are emitted during squark/gluino
decay cascades.
In a truly ``worst case scenario'' winos, higgsinos, sleptons, stops and
sbottoms are not produced neither in squark nor in gluino decays.
Bino $\to Z+$ singlino decays are impossible, if the bino mass is below
$M_Z$. In Table~1 we give the parameters and particle masses
of a benchmark point with these
properties, for which physical masses and decay branching fractions
have been obtained with the public code {\sf NMSSMTools\_4.2.1}
\cite{Ellwanger:2004xm,Ellwanger:2005dv}.

\begin{table}[h!]
\begin{center}
\begin{tabular}{|c|c||c|c||c|c|}
\hline
Parameter & Value & Parameter & Value & Particle(s) & Mass \\
\hline
%UE
$\lambda$ & $6.5\times 10^{-3}$ & $M_1$ & 90 GeV & $M_{H_1}$ & 83 GeV \\
\hline
$\kappa$ & $1.9\times 10^{-5}$ & $M_2$ & 950 GeV & $M_{H_2}$ & 123.2 GeV\\
\hline
$\tan\beta$ & $20$ & $M_3$ & 830 GeV & $M_{H_3,A_2,H^\pm}$ &$\sim 950$ GeV \\
\hline
$\mu_\mathrm{eff}$ & $900$ GeV & $A_t$ & -1500 GeV & $M_{A_1}$ & 12.9 GeV \\
\hline
$\xi_S$ & $-1.02\times 10^{9}$ GeV$^3$ & $A_b$ & -1000 GeV & $M_\mathrm{squarks}$ &
$\sim 860$ GeV \\
\hline
${m'_S}^2$ & $3.6\times 10^{3}$ GeV$^2$ & $m_\mathrm{sleptons}$ & 600 GeV & 
$M_\mathrm{stop1}$ & 810 GeV \\
\hline
$A_\kappa$ & 0 GeV & $m_\mathrm{squarks}$ (u,d,s,c) & 830 GeV & 
$M_\mathrm{stop2}$ & 1060 GeV \\
\hline
$A_\lambda$ & 50 GeV & $m_\mathrm{squarks}$ (t,b) & 900 GeV & 
$M_\mathrm{gluino}$ & 893 GeV \\
\hline
& & & & $M_{\chi^0_1}$ & 5.26 GeV\\
%UEend
\hline
& & & & $M_{\chi^0_2}$ & 89 GeV\\
\hline
\end{tabular}
\caption{Parameters (left and middle column) and particle masses (right column)
of a NMSSM benchmark point.}
\end{center}
\end{table}

The parameters $M_1,\ M_2,\ M_3\ A_t,\ A_b$ in Table~1 denote
the soft SUSY breaking bino-, wino- and gluino mass terms and
Higgs-stop, Higgs-sbottom trilinear couplings, respectively. The
lightest neutralinos are denoted by $\chi_1^0$ (singlino-like)
and $\chi_2^0$ (bino-like), respectively. The
gaugino mass terms are non-universal, but lead to a go-theorem
for branching fractions corresponding to a simplified model:
Squarks decay to 100\% into $\chi_2^0$ and the corresponding quark,
$\chi_2^0$ to 100\% into $\chi_1^0 + H_1$. Gluinos decay with approximately
equal branching fractions only into squarks + quarks of the first two generations.
Hence the decay chains are
\bea\label{eq:2.5}
\tilde{q} &\to& q\, \chi_2^0 \to q\, H_1\, \chi_1^0\; ;\nn \\
\tilde{g} &\to& \tilde{q}\, q\; .
\eea

The small value of $\lambda$ suppresses mixings of the singlet-like $A_1$ with the
doublet-like $A_2$ such that $\chi_2^0$ decays into $\chi_1^0 + H_1$ in spite
%UE
of the lighter $A_1$, and suppresses mixings of the singlet-like $\chi_1^0$
with higgsinos/winos which can lead to unacceptable invisible decay rates of
the Standard Model-like $H_2$ into $\chi_1^0+\chi_1^0$.
%UEend
Despite the small value of $\lambda$ (but due to
the large value of $\tan\beta$) the mass of  $H_2$
-- still consistent with $\sim 126$~GeV within the expected theoretical
error in {\sf NMSSMTools} -- is larger than in the MSSM due to mixing of
$H_2$ with $H_1$ \cite{Badziak:2013bda}. All phenomenological
constraints (except for the muon anomalous magnetic moment) tested
in {\sf NMSSMTools} are satisfied. Due to the small width of $A_1$ it
is difficult, however, to determine the dark matter relic density
accurately with the code {\sf micrOMEGAs} \cite{Belanger:2013oya} inside
{\sf NMSSMTools} -- for our benchmark point its numerical value seems smaller than
the desired value $\Omega h^2 \sim 0.12$ \cite{Hinshaw:2012aka,Ade:2013zuv},
which shows in any case that the relic density can be reduced sufficiently.

We have simulated events at the LHC at 8~TeV for this point using
MadGraph/\-MadEvent~\cite{Alwall:2011uj} which
includes Pythia~6.4 \cite{Sjostrand:2006za} for showering and
hadronisation. The emission of one additional hard jet was allowed in the
simulation; the production cross sections for squark-squark, squark-gluino,
squark-antisquark and gluino-gluino production were obtained by
Prospino at NLO \cite{Beenakker:1996ch,Beenakker:1996ed}, including
correction factors from the resummation of soft gluon emmission
estimated from \cite{Beenakker:2011fu,Kramer:2012bx}. 
(At 8~TeV, the total squark-gluino production cross section for the
%UE
benchmark point is $\sim$524~fb.) The output in
%UEend
StdHEP format was given to CheckMATE \cite{Drees:2013wra} which includes
the detector simulation DELPHES \cite{deFavereau:2013fsa} and compares
the signal rates to constraints in various search channels of ATLAS and
CMS. Additional analyses were performed by means of MadAnalysis~5
\cite{Conte:2012fm,Conte:2013mea}.

Following CheckMATE, the signal rates for the above benchmark point
are compatible with constraints from available search channels for SUSY. In spite of the
many $b$-jets from $H_1$ decays, the dominant constraints for this point
originate from the search for jets and $E_T^\mathrm{miss}$ in
\cite{TheATLAScollaboration:2013fha}, more precisely from channel~D
requiring 5~hard jets, $E_T^\mathrm{miss} > 160$~GeV and
$E_T^\mathrm{miss}/m_\mathrm{eff}(5j) > 0.2$, where $m_\mathrm{eff}(Nj)$ is the
scalar sum of transverse momenta of the leading N jets and
$E_T^\mathrm{miss}$.

Searches for events at the LHC with jets and $b\bar{b}$ pairs have also been
performed by ATLAS in~\cite{ATLAS-CONF-2014-005} however aiming at resonances
in the $b\bar{b}b\bar{b}$ final state. 
%UE
Also the upper bounds on signal rates
in $b\bar{b}\gamma\gamma$ final states from CMS~\cite{CMS-PAS-HIG-13-032}
and ATLAS~\cite{Aad:2014yja} are satisfied, amongst others since the $\gamma\gamma$
invariant mass required there does not cover our range of $M_{H_1}$.
%UEend

An $M_{T2}$ Higgs analysis was performed by CMS in \cite{CMS-PAS-SUS-13-019},
which aimed at a scenario similar to those discussed here: squark/gluino decay
cascades ending in $\chi_2^0 \to \chi_1^0 + H_{SM}$. In one of the
considered channels (high $H_T$ region) the cut on $E_T^\mathrm{miss}$
was lowered to $E_T^\mathrm{miss} > 30$~GeV. The absence of significant
excesses was interpreted in terms of a gluino-induced simplified model
leading to $M_\mathrm{gluino} \gsim 850$~GeV (depending in $M_{\chi_1^0}$);
however, $M_{\chi_2^0}=M_{\chi_1^0}+200$~GeV was assumed in this analysis.

%UE
Further searches for excesses in events with jets without lower cuts on
$E_T^\mathrm{miss}$ have been performed in \cite{ATLAS-CONF-2013-091,
CMS-PAS-EXO-12-049,Chatrchyan:2013izb,Chatrchyan:2013xva}. The most
constraining search channel in \cite{ATLAS-CONF-2013-091} is the one
requiring 7~jets with $p_T > 80$~GeV and at least two tagged $b$-jets.
After a simulation we find about 240 events in this channel for the
benchmark point, complying with the data at the 2~$\sigma$ level.
Concerning the search for three-jet resonances in \cite{CMS-PAS-EXO-12-049},
we find that the two $b$-jets from $H_1$ decays merge sufficiently often
into one single jet such that the event rates and acceptance are about
20~times smaller than the one assumed in \cite{CMS-PAS-EXO-12-049} for
gluino pair production with RPV decays into three jets, and the limits
are well satisfied. Regarding the search for two-jet resonances in
\cite{Chatrchyan:2013izb} we find that the average two-jet mass peaks
at $\sim 800$~GeV (somewhat below the squark/gluino masses) and the
acceptance after cuts (within a window of a width of $\sim 15\%$ times
the mass) is $\lsim 4\%$; consequently the signal complies with the limits
shown in Fig.~3 in \cite{Chatrchyan:2013izb}. Finally we have studied
$S_T$, the sum of $|p_T|$ of objects with $|p_T| > 50$~GeV relevant for
the search for microscopic black holes in \cite{Chatrchyan:2013xva}.
For all multiplicities $N \gsim 3...10$ the signal events are below
$10\%$ of the data points shown in Figs.~2 and Figs.~3 in
\cite{Chatrchyan:2013xva} without any peak-like structure, and thus this
search is not restrictive.
%UEend
Hence present analyses, potentially sensitive to the decay products of
two $H_1$ bosons in the final state, are not sensitive to the
benchmark point.

\begin{figure}[t!]
\begin{center}
\psfig{file=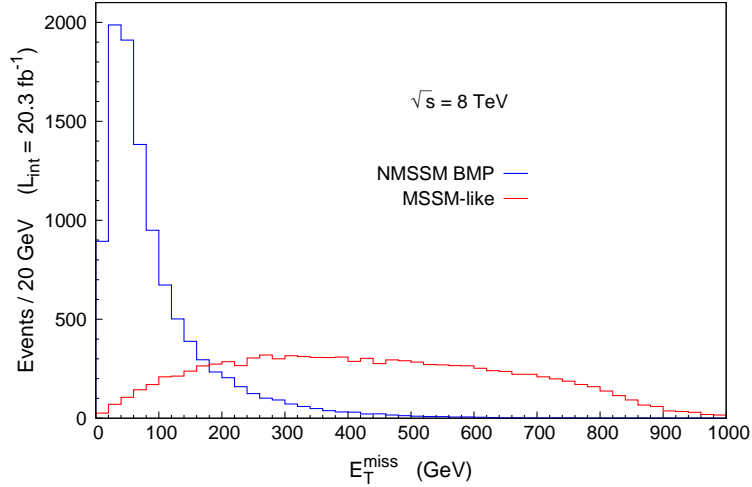, scale=0.8}
\end{center}
\vspace*{-8mm}
\caption{Spectra of $E_T^\mathrm{miss}$ before cuts for the
benchmark point (blue) and a similar point in the MSSM with a bino LSP
(red).}
\label{fig:1}
\end{figure}

Given that the masses of the gluino and the squarks of
the first generation are well below 1~TeV it is clear that a corresponding point in the
parameter space of the MSSM would be well excluded, the reason being
the different spectra of $E_T^\mathrm{miss}$. To clarify this effect
we show in Fig.~\ref{fig:1} the spectrum of $E_T^\mathrm{miss}$ for the
benchmark point and for a similar point in the MSSM, which differs from
the benchmark point only in a stable bino, which is now the LSP.

In Fig.~\ref{fig:1} one sees the dramatic reduction of
$E_T^\mathrm{miss}$ due to the
NLSP $\to$ LSP + $H_1$ decay; the few remaining
events with large $E_T^\mathrm{miss}$ for the benchmark point (denoted by
NMSSM~BMP in Fig.~\ref{fig:1}) originate from neutrinos from $\tau$ and $b$
decays after $H_1 \to \tau^+\tau^-,\ b\bar{b}$ 
(and, to a minor extent, from the singlino).

In fact, the final states from $H_1$ decays not only reduce $E_T^\mathrm{miss}$,
but lead also to an increase of $m_\mathrm{eff}(Nj)$. Hence the cut
$E_T^\mathrm{miss}/m_\mathrm{eff}(5j) > 0.2$ in
\cite{TheATLAScollaboration:2013fha} reduces the number of signal
events even more dramatically by a factor $\sim 0.07$,
and events passing this cut satisfy
$E_T^\mathrm{miss} > 160$~GeV automatically. (For the MSSM-like point,
channel~D is actually not the most constraining channel, but rather
channel AM.)

How sensitive are these reductions of signal events to the masses of the
involved particles? In order to answer this question we have varied both
the singlino mass $M_{\tilde{S}}$ from 1~GeV to 17~GeV and $M_{H_1}$
from 87~GeV to 71~GeV, keeping the bino
mass fixed at 89~GeV. We first studied the ratio $R^{E_T^\mathrm{miss}}$
defined by the ratio
of the number of events with $E_T^\mathrm{miss} > 160$~GeV (before other
cuts) in the NMSSM, over the number of events in the MSSM with the bino
as LSP. The results for $R^{E_T^\mathrm{miss}}$ are shown in Table~2.
(The relative statistical error on $R^{E_T^\mathrm{miss}}$ is about 2\%
for $R^{E_T^\mathrm{miss}} \sim 0.15$, decreasing slightly with increasing
$R^{E_T^\mathrm{miss}}$.)

%\begin{center}
\begin{table}[t!]
\begin{center}
\begin{tabular}{|c||c|c|c|c|c|c|c|c|c|}
\hline
$M_{\tilde{S}}$ (GeV):&1&3&5&7&9&11&13&15&17\\
\hline
$M_{H_1}$ (GeV): & \multicolumn{9}{c|} {$R^{E_T^\mathrm{miss}}$: }\\
\hline
87&	.125&	    &       &       & 	    &	    &	    &	    &	    \\\hline
85&	.134&	.134&	    &	    &	    &	    &	    &	    &	    \\\hline
83&	.147&	.146&	.145&	    &	    &	    &	    &	    &	    \\\hline
81&	.166&	.169&	.161&	.160&	    &	    &	    &	    &	    \\\hline
79&	.192&	.194&	.186&	.186&	.179&	    &	    &	    &	    \\\hline
77&	.232&	.224&	.225&	.221&	.211&	.207&	    &	    &	    \\\hline
75&	.273&	.276&	.268&	.261&	.266&	.252&	.247&	    &	    \\\hline
73&	.319&	.316&	.309&	.310&	.307&	.302&	.298&	.294&	    \\\hline
71&	.358&	.366&	.362&	.359&	.353&	.355&	.353&	.345&	.343\\\hline
\end{tabular}
\caption{Ratios $R^{E_T^\mathrm{miss}}$
of the number of events with $E_T^\mathrm{miss} > 160$~GeV (before other
cuts) in the NMSSM, over the number of events in the MSSM (with the bino
as LSP), as function of $M_{\tilde{S}}$ and $M_{H_1}$ keeping $M_{\mathrm{bino}}$
fixed at 89~GeV.}
\end{center}
\end{table}
%\end{center}

We see that, for the singlino mass $M_{\tilde{S}}$ in the kinematically
allowed range, $R^{E_T^\mathrm{miss}}$ varies little with $M_{\tilde{S}}$
for fixed $M_{H_1}$:
on average, $R^{E_T^\mathrm{miss}}$ decreases slightly with increasing
$M_{\tilde{S}}$ towards the
boundary of phase space. On the other hand, for fixed $M_{\tilde{S}}$,
$R^{E_T^\mathrm{miss}}$ has a stronger increase with decreasing $M_{H_1}$
(away from the boundary of phase space).

\begin{figure}[t!]
\begin{center}
\psfig{file=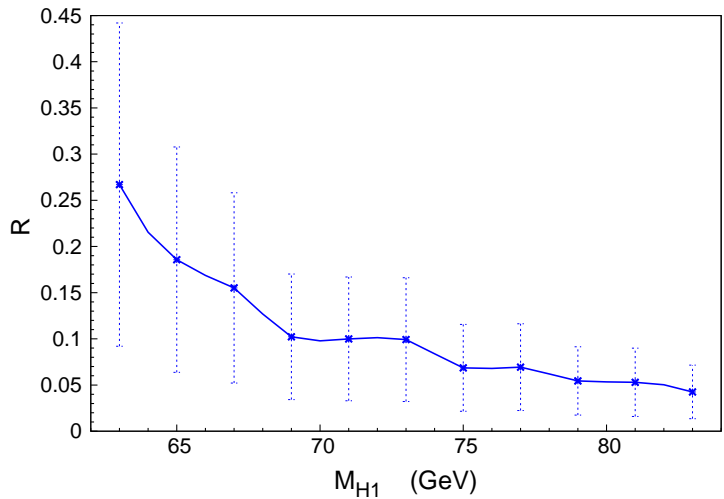, scale=0.8}
\end{center}
\vspace*{-8mm}
\caption{The ratio $R$ of
signal events in the NMSSM over the number of events in the MSSM with the bino
as LSP, after all cuts for channel~D in \cite{TheATLAScollaboration:2013fha},
as function of $M_{H_1}$ keeping the singlino mass fixed at 5.3~GeV.}
\label{fig:2}
\end{figure}

As stated above, the impact of the ``missing'' $E_T^\mathrm{miss}$ on the
signal rates in channel~D~\cite{TheATLAScollaboration:2013fha}
after all cuts including $E_T^\mathrm{miss}/m_\mathrm{eff}(5j)
> 0.2$ is actually stronger. In Fig.~\ref{fig:2} we show the ratio of
signal events in the NMSSM over the number of events in the MSSM with the bino
as LSP, after all cuts for this channel, as function of $M_{H_1}$ (keeping
the singlino mass fixed at the value of the benchmark point of 5.3~GeV).
The error bars indicate the statistical fluctuations from our simulations
as determined by CheckMATE.

We see that the reduction of signal events remains very strong, even if
$M_{H_1}$ is several GeV away from the boundary of phase space. Hence the
result of the analyses summarised in Table~1 and Fig.~\ref{fig:2} is that
suppressing the number of signal events in typical SUSY search channels
does not require a particular fine-tuning of masses.

On the other hand we should keep in mind that contributions from
neutrinos from squark decay cascades to $E_T^\mathrm{miss}$ have been
suppressed for all points considered above; still it seems important to
re-interpret the absence of excesses in typical SUSY search channels
in terms of lower bounds on squark/gluino masses for such configurations
of bino-, singlino- and $H_1$ masses.

The next question is whether other search strategies, not relying on strong
cuts on $E_T^\mathrm{miss}$, can be sensitive to such difficult SUSY scenarios with
two $H_1$ bosons in the final state. A first proposal towards extracting corresponding
signals -- which will be improved in a future publication -- is presented
in the next section.

\section{Towards the extraction of signals in $b\bar{b}+\tau^+\tau^-$
 final states at 14~TeV}

Signals from Higgs production via neutralino decays in sparticle decay
cascades have been analysed previously, mostly in the context of the MSSM
in \cite{Datta:2003iz,Bandyopadhyay:2008fp,Huitu:2008sa,
Bandyopadhyay:2008sd,Gori:2011hj,Stal:2011cz,Kribs:2009yh,Kribs:2010hp,
Belyaev:2012si,Bhattacherjee:2012bu,Belyaev:2012jz}. 
There, however, significant lower cuts on
$E_T^\mathrm{miss}$ were applied, since in the considered scenarios the
LSP had no reason to be particularly soft.

In the present scenario the final states of squark and gluino production
are characterised by jets with large $p_T$, little $E_T^\mathrm{miss}$,
but remnants of two $H_1$ Higgs bosons whose branching ratios coincide essentially
with those of a Standard Model-like Higgs boson of the corresponding
mass. (The prospects for a direct discovery of $H_1$ at the LHC are rather
dim: its couplings to SM particles squared, and hence its production
cross sections and signal rates, are only $6\%-6.5\%$ of the ones of a
SM-like Higgs boson of similar mass.)

Subsequently we describe an approach towards the extraction of a possible
signal in the $b\bar{b}+\tau^+\tau^-$ final state, using the same
simulation methods described in the previous section.

\begin{figure}[ht!]
\begin{center}
\psfig{file=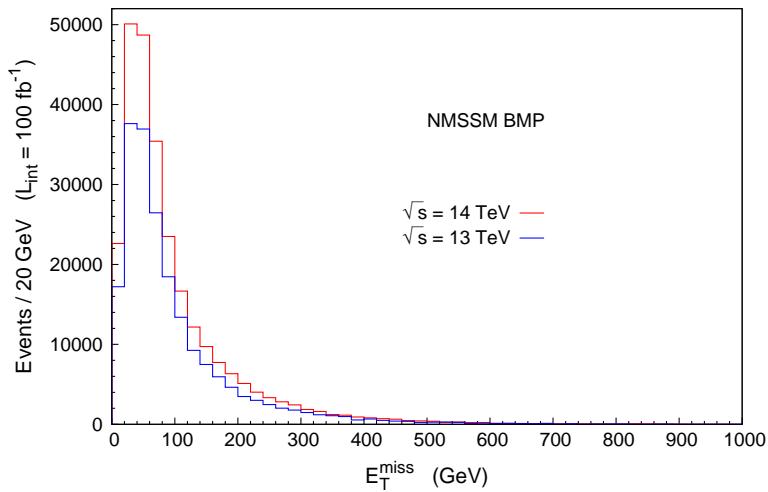, scale=0.8}
\end{center}
\vspace*{-8mm}
\caption{Expected $E_T^\mathrm{miss}$ distribution for the benchmark point
at the LHC at 13 and 14~TeV c.m. energy.}
\label{fig:3}
\end{figure}

First we have studied $E_T^\mathrm{miss}$ for the benchmark
point for $pp$ collisions at the LHC at 13~TeV and 14~TeV
c.m. energy assuming an integrated luminosity of 100~fb$^{-1}$. As can
be seen in Fig.~\ref{fig:3},
$E_T^\mathrm{miss}$ is still peaked at low values.

\begin{figure}[t!]
\begin{center}
\psfig{file=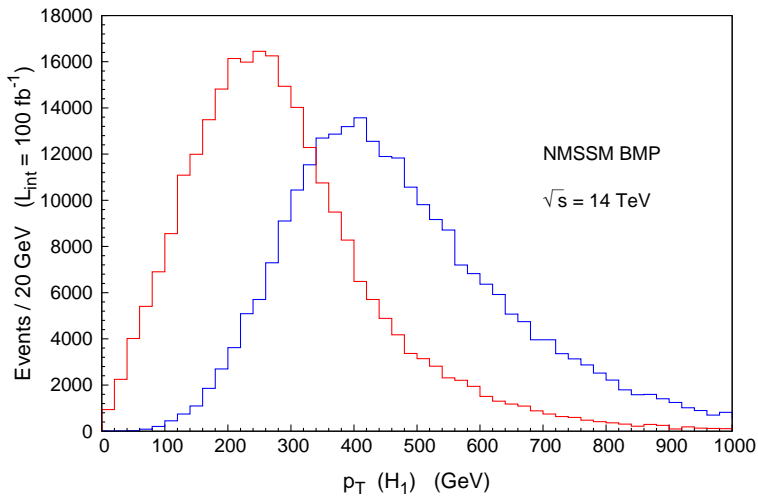, scale=0.8}
\end{center}
\vspace*{-8mm}
\caption{Transverse momentum distributions of the leading $H_1$ (blue)
and next-to-leading $H_1$ (red) after squark and gluino production
at the LHC at 14~TeV c.m. energy for the benchmark point.}
\label{fig:4}
\end{figure}

When looking for the remnants of two $H_1$ Higgs bosons,
an important question is
which transverse momenta can one expect for these particles. In
Fig.~\ref{fig:4} we show the leading and next-to-leading $p_T$
distribution corresponding to the benchmark point.
(Here and in the following we concentrate on 14~TeV c.m. energy.)
We see that the $p_T$ of the leading $H_1$ is peaked near 400~GeV,
and the $p_T$ of the next-to-leading $H_1$ is peaked near 200~GeV,
which can be used for cuts on the final states.

We next describe the sequence of cuts which were applied. The analysis
of the events was performed in two steps: to start with, jets were
constructed by Fastjet \cite{Cacciari:2011ma} (part of the Delphes package
\cite{deFavereau:2013fsa}) using the anti-$k_T$ algorithm
\cite{Cacciari:2008gp} and a jet cone radius $R=0.5$. This value was chosen
such that, as often as possible, all decay products of the leading
$H_1$ (but no other hadrons) are part of a single jet whose mass distribution is
analysed at the end. 

We require four hard jets (including $b$-tagged jets) with $p_T > 400\ \mathrm{GeV},\
200\ \mathrm{GeV},\ 80\ \mathrm{GeV},$ $80\ \mathrm{GeV}$,
respectively. A significant $E_T^\mathrm{miss}$ is not part of the
signal; some $E_T^\mathrm{miss}$ can be expected, however, from
neutrinos of $\tau$ decays once we require $2\ \tau$ in the
final state (see Fig.~\ref{fig:3}).
Hence we only impose $E_T^\mathrm{miss} > 30$~GeV.

Then, jets of the same event were reconstructed with a
jet cone radius $R=0.15$. The aim is to identify as many ``slim''
$b$-jets as possible, together with their kinematic properties.
The value $R=0.15$ is just marginally larger than the granularity
$0.1\times 0.1$
of the ATLAS hadronic tile calorimeter (we use the ATLAS detector card inside
Delphes). For $b$-tagged jets we require $p_T > 40$~GeV and assume a
$b$-tag efficiency of 70\% (mistag efficiencies from $c$-jets
of 10\%, and from light quark/gluon jets of 1\%).

Among the jets reconstructed with $R=0.15$ we require $\geq 2$
$b$-jets and $\geq 2$ hadronic $\tau$ leptons. 
Since the invariant mass of the pair of $\tau$-leptons is difficult
to reconstruct we just require $M_{\tau\tau} < 120$~GeV and, in order to
remove the background from fake $\tau$ leptons (see below), $M_{\tau\tau} > 20$~GeV.
The 2~$b$-jets next to each
other (with the smallest $\Delta R$) are combined into a
$2b$-pseudojet, 2bPJ.
Among the jets constructed with a jet cone radius $R=0.5$ we look
for the one closest to the 2bPJ (with $\Delta R < 0.1$); this
jet $\widehat{J}$
is our candidate for the remnants of $H_1 \to b\bar{b}$.
For $\widehat{J}$ we further require $p_T > 400$~GeV.  The
invariant mass of the 2bPJ should be smaller than the mass of
$\widehat{J}$. (The mass of the 2bPJ can be considerably smaller due
to radiation off the $b$-quarks not included in the $R=0.15$ jets.)

Finally we require the mass of $\widehat{J}$ 
to be 40~GeV~$< M_{\widehat{J}} <$~120~GeV,  
and plot $M_{\widehat{J}}$ in this range.

%\begin{center}
%UE
\begin{table}[h!]
\begin{tabular}{|c|c|c|}
\hline
 & Benchmark point & $jjb\bar{b}$ background \\
\hline
Cross section in fb & 5232 & $1.47\times 10^{5}$\\
\hline
$p_T$(jets) $> 400,200,80,80$~GeV & 3513& 6835 \\
\hline
$E_T^\mathrm{miss} > 30$~GeV & 3118& 3875 \\
\hline
$\geq 2\ b$-jets, $\geq 2\ \tau_h$ with 
$20\ \mathrm{GeV} < M_{\tau\tau}< 120$~GeV & 99.3& 97.7 \\
\hline
$\Delta R (\widehat{J}, \mathrm{2bPJ}) < 0.1$& 48.9& 37.8\\
\hline
$p_T(\widehat{J}) > 400$~GeV& 27.1& 12.9\\
%\hline
%$\Delta \varphi(E_T^\mathrm{miss},\widehat{J}) > 0.3$ & 31.3& 11.7\\
\hline
$M_\mathrm{2bPJ} < M_{\widehat{J}}$ & 24.1&  9.6\\
\hline
$40\ \mathrm{GeV} < M_{\widehat{J}} < 120$~GeV & 22.8& 7.5\\
\hline
\end{tabular}
\caption{Impact of the cuts described in the text on the
event rates of the benchmark point and the dominant $jjb\bar{b}$ background
(the latter after cuts at the parton level as described in the
text).}
\end{table}
%\end{center}

\begin{figure}[ht!]
\begin{center}
\psfig{file=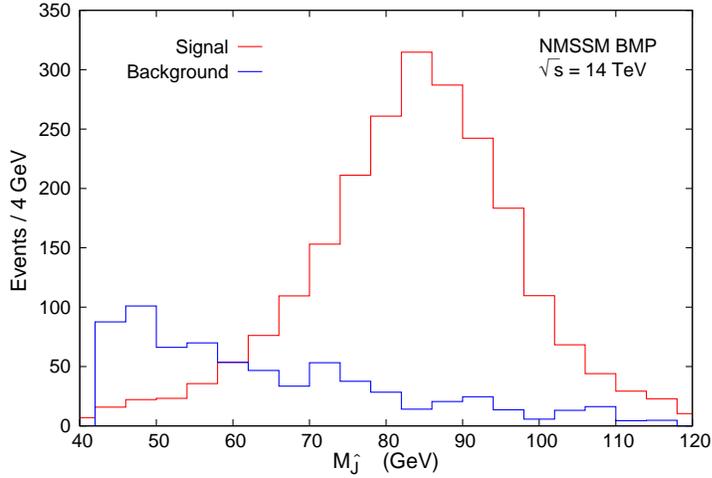, scale=0.8}
\end{center}
\vspace*{-8mm}
\caption{Plot of $M_{\widehat{J}}$ (in red) for the benchmark point at 14~TeV c.m.
energy, after application of the cuts described in the text and in Table~3.
The background contribution from $jjb\bar{b}$ with two mistagged $\tau$
leptons is shown in blue.}
\label{fig:5}
\end{figure}

The result displayed in Fig.~\ref{fig:5}, which is based on the simulation of
$\sim 230 000$ events, shows that $M_{\widehat{J}}$ peaks indeed near
%UEend
the mass of $H_1$, 83~GeV for the benchmark point considered here.
The event rates are normalised to an integrated luminosity of
100~fb$^{-1}$; for the total squark-gluino production cross section at 14~TeV 
we have 5232~fb at NLO+NNLO. The impact of the above cuts is shown
in Table~3; within the signal region 40~GeV$ < M_{\widehat{J}} < 120$~GeV
the cross section is about 23~fb.
Given the $H_1 \to \tau^+\tau^-$ branching fraction of
about 8\% and the tagging efficiencies, the dominant reduction of
signal events results from the requested $\geq 2$ $b$-jets and
$\geq 2\ \tau$ leptons.

A priori, the following Standard Model backgrounds contribute to the
final states defined above: $jjb\bar{b}\tau^+\tau^-$ from
QCD and electroweak production, $jjt\bar{t}$ possibly with additional
mistagged $\tau$ leptons, and $jjb\bar{b}$ with two mistagged $\tau$
leptons. We have estimated these backgrounds using the same simulation
methods applied for the signal. However, in order to make our cut analysis
sufficiently efficient, we also applied cuts at the parton level in
MadGraph on jets (quarks and gluons) and $b$-quarks: For $b$-quarks
we required $p_T > 100$~GeV and $|\eta| < 2$, and for the four leading
jets (including $b$-quarks) $p_T > 200~\mathrm{GeV}, 100~\mathrm{GeV},
80~\mathrm{GeV}, 80$~GeV, respectively.
We checked that these cuts do not generate a bias in the $M_{\widehat{J}}$
spectrum after applying the additional cuts described above and in Table~3.

After the cuts, the background from $jjb\bar{b}\tau^+\tau^-$ from
QCD and electroweak production turned out to be negligibly small, with a
cross section in the signal region of about $0.007$~fb. 
After all cuts, the background from $jjt\bar{t}$ is also small, with a
cross section in the signal region of about $0.44$~fb.

However, the background from $jjb\bar{b}$ with two mistagged $\tau$
leptons is relatively large due to the fact that we tag for $b$-jets
and $\tau$-leptons using jet reconstruction with a small jet cone radius
$R=0.15$. Many of such ``slim'' jets are mistagged as $\tau$-leptons,
often as pairs with a relatively low invariant mass. A priori, about
5\% of all $jjb\bar{b}$ events after cuts contain such a fake $\tau$
pair. This fake rate can be reduced by a factor $\sim \frac{1}{2}$
after a cut $M_{\tau\tau} > 20$~GeV, which reduces the signal by only
about 12\%. Then this background results in a
cross section in the signal region of about $7$~fb. Its contribution
to $M_{\widehat{J}}$, based on the simulation of 300000 events,
is also shown in Fig.~\ref{fig:5}, and it seems that the signal can
be distinguished clearly from this background.

In practise the background is often
obtained from data-driven control regions. Once it is measured,
modifications of the
cuts given above and/or additional cuts (for example on the absence of
isolated leptons) are likely to improve the signal/background ratio
even further.

\begin{figure}[b!]
\begin{center}
\psfig{file=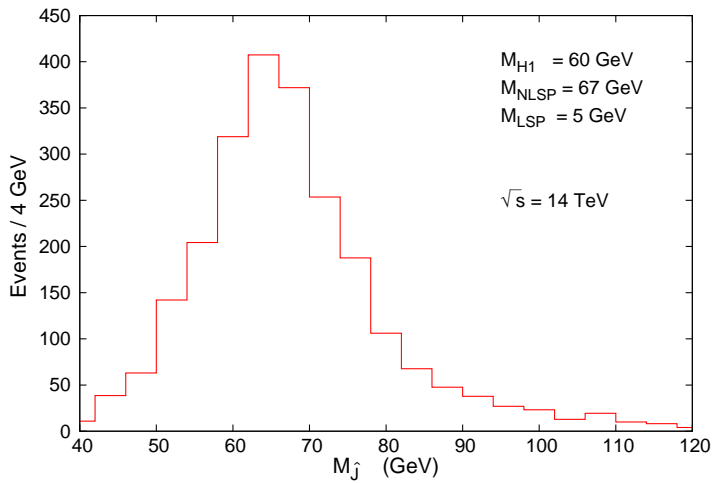, scale=0.8}
\end{center}
\vspace*{-8mm}
\caption{Plot of $M_{\widehat{J}}$ at 14~TeV c.m.
energy, for a point with $M_{H_1}=60$~GeV, $M_\text{NLSP} = 67$~GeV and
$M_\text{LSP} = 5$~GeV
after application of the cuts described in the text and in Table~3.}
\label{fig:6}
\end{figure}

Of course, the mass $M_{H_1}$ can differ from the value $M_{H_1}=$~83~GeV 
assumed for the benchmark point. In order to see the impact of a lighter
$H_1$ we have repeated the simulation for a point with the same
sparticle spectrum but $M_{H_1}=60$~GeV, $M_\text{NLSP} = 67$~GeV and
$M_\text{LSP} = 5$~GeV. The resulting $M_{\widehat{J}}$ spectrum is shown
in Fig.~\ref{fig:6} and again we see that the sequence of cuts allows, in
principle, to identify $M_{H_1}$ from the $M_{\widehat{J}}$ spectrum.

The $H_1 H_1$ final state, characteristic of squark/gluino production
in the present scenario, resembles actually Higgs pair production in
the Standard Model up to the unknown $H_1$ mass but, 
provided that squarks and gluinos are not excessively heavy, with an
associated larger cross section 
(see
\cite{Dolan:2012rv,Papaefstathiou:2012qe, Barr:2013tda, Barger:2013jfa,
deLima:2014dta} for some recent studies in the Standard Model).
Corresponding techniques like more refined subjet-based analyses applied to
the $b\bar{b}\,\tau^+\tau^-$ final state in \cite{Dolan:2012rv} can
probably be useful here as well.

\begin{figure}[t!]
\begin{center}
\psfig{file=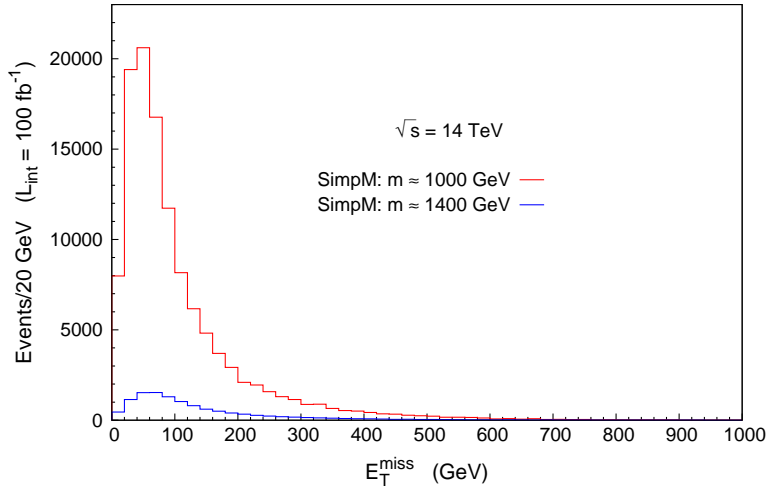, scale=0.8}
\end{center}
\vspace*{-8mm}
\caption{Expected $E_T^\mathrm{miss}$ distribution
at the LHC at 14~TeV c.m. energy for two simplified models with
squark $\sim$ gluino masses of 1000~GeV and 1400~GeV.}
\label{fig:7}
\end{figure}

The squark/gluino production cross sections would decrease, of course,
for squarks and/or gluinos heavier than their benchmark value
(860/890~GeV, respectively), and the kinematics will change. 
First we investigate the impact of heavier squarks and gluinos on
the shape of $E_T^\mathrm{miss}$. 
We illustrate this with simplified models where, 
as in the case of the previous benchmark point, squarks
decay with a 100\% branching ratio into the bino-like NLSP
(still with a mass of 89~GeV) which can only decay (with 100\% BR) 
into $H_1$ and the singlino-like LSP (both still with masses of 83~GeV and 5~GeV,
respectively).
The corresponding $E_T^\mathrm{miss}$ distributions are shown in Fig.~\ref{fig:7}
for squark/gluino masses of 1000~GeV or 1400~GeV.
(The gluinos are taken 5~GeV heavier than squarks to allow, for simplicity, for
flavour democratic gluino 3-body decays into quarks + squarks of the first
two generations.)

\begin{figure}[t!]
\begin{center}
\psfig{file=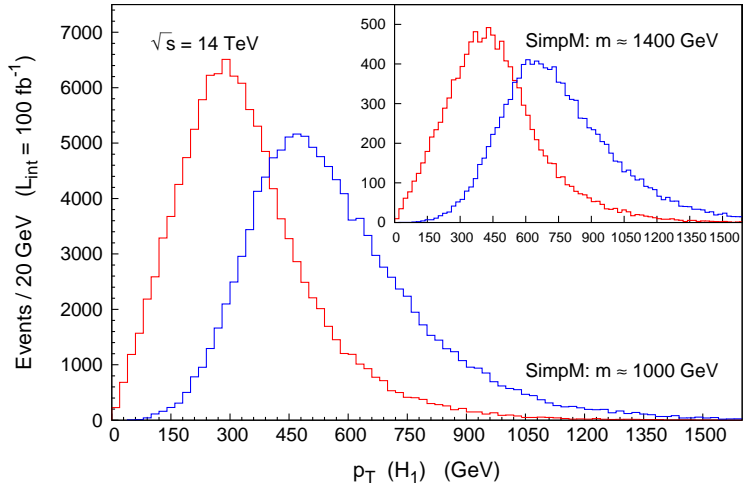, scale=0.8}
\end{center}
\vspace*{-8mm}
\caption{Transverse momentum distributions of the leading $H_1$ (blue)
and next-to-leading $H_1$ (red) after squark and gluino production
at the LHC at 14~TeV c.m. energy. We assume simplified spectra, with 
squarks/gluino masses of
1000~GeV and 1400~GeV.}
\label{fig:8}
\end{figure}

One finds that $E_T^\mathrm{miss}$ still strongly peakes at low
values; hard cuts on $E_T^\mathrm{miss}$ would remove again most of the
signal events.
The transverse momenta of the the leading
and next-to-leading $H_1$ for squarks/gluinos with masses of
1000~GeV and 1400~GeV, respectively, are shown
in Fig.~\ref{fig:8}.

\begin{figure}[t!]
\begin{center}
\psfig{file=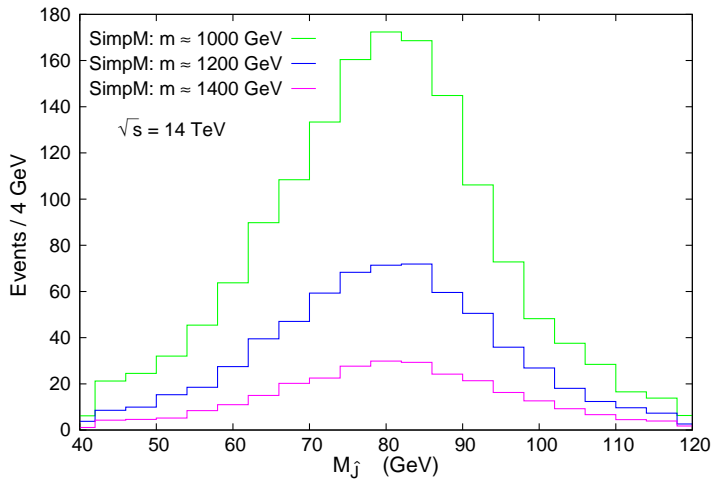, scale=0.8}
\end{center}
\vspace*{-8mm}
\caption{$M_{\widehat{J}}$ at 14~TeV c.m.
energy for simplified model spectra with $M_{H_1}=83$~GeV, $M_\text{NLSP} = 89$~GeV and
$M_\text{LSP} = 5$~GeV, and squark/gluino masses of 1000~GeV,
1200~GeV and 1400~GeV, respectively.}
\label{fig:9}
\end{figure}

As visible from Fig.~\ref{fig:8}, the average transverse momenta of the $H_1$ states are
considerably larger for heavier squarks/gluinos.
Finally we ask whether the shape of the
$M_{\widehat{J}}$ spectrum changes for heavier squarks/gluinos.
Using the same analysis and cuts as before, the resulting 
$M_{\widehat{J}}$ spectrum is
shown in Fig.~\ref{fig:9} for squark/gluino masses of 1000~GeV,
1200~GeV and 1400~GeV.
We see that the shape of the $M_{\widehat{J}}$ spectrum remains
unchanged; only the signal rate decreases as expected (slightly less,
in fact) as does the production cross section which is now 
about 2226~fb,
693~fb and 242~fb for squark/gluino masses of 1000~GeV,
1200~GeV and 1400~GeV, respectively. However, for squark/gluino
masses above 1400~GeV, the signal obtained with the
present cuts (and jet analysis) starts to fall below the background from
$jjb\bar{b}$ with two mistagged $\tau$ leptons (shown in Fig.~\ref{fig:5}).

On the other hand, since the production of heavier squarks and gluinos
generates both $H_1$ states and jets with larger transverse momenta, cuts can
be optimised. Search channels with significantly harder cuts on the
transverse momenta of at least the candidate $\widehat{J}$-jets
(assumed to contain the remnants of the leading $H_1$) can be employed.
Corresponding analyses will be the subject of future publications.

\section{Conclusions and outlook}

The most important result of the present paper is the existence of
scenarios in the general NMSSM in which a light singlino at the end
of sparticle decay cascades reduces strongly the missing transverse
energy, one of the essential criteria in standard search channels
for supersymmetry. In such ``worst case scenarios'' hardly any missing
transverse energy is produced along each step of sparticle decay cascades.
We present a realistic benchmark point, consistent with the properties of the
Standard Model-like Higgs boson at $\sim 126$~GeV and the dark matter
relic density, satisfying present constraints from SUSY search channels
with all sparticle masses below $\sim 1$~TeV.

The two NMSSM-like Higgs bosons $H_1$, produced in each event of squark,
squark-gluino or gluino pair production in this scenario, allow for new
search channels which do not rely on large missing transverse
energy, but on the $H_1$ decay products. We have presented an analysis
which shows that, for not too heavy squarks/gluinos, a $H_1$ signal
can be visible above the Standard Model background in the $b\bar{b}\tau^+\tau^-$
+ jets final state. This analysis can certainly be improved in various aspects,
but already indicates the lines along which a signal can be obtained.

Among the possible improvements are analyses based on jet substructure as
is the case of Higgs pair production into the same final state in \cite{Dolan:2012rv}
(replacing the step of our analysis based on a jet cone radius of 0.15),
which may also help to reduce the background from mistagged tau pairs.
Also searches for a $(b\bar{b})(b\bar{b})$ final state as in
\cite{deLima:2014dta} might be feasible. Finally, in order to get some
direct information on the masses of the originally produced squark/gluino
pairs (beyond the production cross section), analyses based on jet
substructure may be combined with analyses based on $M_{T2}$ as, for
example, in \cite{CMS-PAS-SUS-13-019}.

Variants of the benchmark scenarios discussed here could also be realised
in principle: First, winos and/or higgsinos could be lighter and appear
in squark decay cascades. Then standard SUSY search channels relying on
$E_T^\mathrm{miss}$ and jets, isolated leptons etc., start to become sensitive to
squark/gluino production but it remains to be studied when, in the
presence of a light singlino and for the kinematical situation discussed
here, such search channels become more sensitive than the type of
analysis presented here.

Second, the final state $"X"$ in the final step NLSP $\to X$ + LSP of
sparticle decay cascades does
not necessarily have to be a light NMSSM-specific Higgs boson $H_1$. Again,
if $X$ is for instance a $Z$ boson or a combination of $Z$ and $H_{\text{SM}}$ bosons
(depending on the branching fractions of the NLSP), standard SUSY search
channels can become relevant since more $E_T^\mathrm{miss}$ is expected
from $X$ decays. However, if the number of events with $E_T^\mathrm{miss}$ is
still reduced due the particular kinematical situation discussed here,
channels which depend less on $E_T^\mathrm{miss}$ but more on $X$
decay products would again be more promising. Such cases merit also to
be studied in the future.

%\vfill

\section*{Acknowledgements}
%UE
We are grateful to Y.~Kats for helpful comments on additional searches
for new physics.
%UEend
We acknowledge help from A. Belyaev for background simulations, and
thank B. Fuks, S. Kulkarni and S. Moretti for helpful discussions.
A.~M.~T. and U.~E. acknowledge support from European Union Initial
Training Network INVISIBLES (PITN-GA-2011-289442). U.~E.
acknowledges support from the ERC advanced grant Higgs@LHC, and from
the European Union Initial Training Network Higgs\-Tools
(PITN-GA-2012-316704).

\end{document}